**Piotr GAS**

AGH University of Science and Technology

# Tissue Temperature Distributions for Different Frequencies derived from Interstitial Microwave Hyperthermia

***Abstract***. *The aim of this study was to evaluate and compare temperature distributions for different tissues being treated at the time of interstitial microwave hyperthermia. A coaxial-slot antenna implemented into the tissue is the source of microwave radiation. The described model takes into account the wave equation for the TM mode and the Pennes equation determining the temperature distribution within the tissue in the stationary case. The simulation results for the three fundamental microwave frequencies of tissue heating devices are presented.*

***Streszczenie***. *Celem pracy było wyznaczenie i porównanie rozkładu temperatury dla różnych tkanek poddawanych leczeniu w czasie śródmiąższowej hipertermii mikrofalowej. Źródłem promieniowania mikrofalowego jest współosiowa antena ze szczeliną powietrzną wprowadzona do wnętrza tkanki. Opisany model uwzględnia równanie falowe dla modu TM oraz równanie Pennesa dla przypadku stacjonarnego określające rozkład temperatury w tkance. Wyniki symulacji zestawiono dla trzech podstawowych częstotliwości pracy urządzeń do grzania mikrofalowego tkanek.* **(Rozkłady temperatury tkanek dla różnych częstotliwości pochodzące z śródmiąższowej hipertermii mikrofalowej)**

**Słowa kluczowe**: śródmiąższowa hipertermia mikrofalowa, fale TM, częstotliwości medyczne, biologiczne równanie ciepła, MES

## Introduction

Interstitial microwave hyperthermia is a kind of thermal therapy which uses high frequency needle electrodes, microwave antennas, ultrasound transducers, laser fibre optic conductors, or ferromagnetic rods, seeds or fluids to treat pathological cells located deep within the human body [1]. The above mentioned elements are directly implanted into diseased tissues and therefore the pathological tissues can be easily heated to a therapeutic temperature of 40 – 46ºC, while surrounding normal tissues are minimally affected [2]. The invasiveness of this method makes potentially the most effective one in successfully used to curing brain, liver, breast, kidney, bone and lung tumors [3]. Heat produced by microwaves can be applied to induce thermonecrosis in tumors and cancerous tissues at the distance of 1 to 2 cm around the heat source. It is worth noting that this technique is suitable for tumors less than 5 cm in diameter [4]. Moreover, microwave hyperthermia is frequently used in conjunction with other cancer therapies, such as radiation therapy or chemotherapy.

Basic recommendations for non-communications applications of the industrial, scientific and medical (ISM) equipments have been designated by the International Telecommunication Union (ITU) [5]. According to ITU, typical ISM devices are designed to generate and utilize locally microwave energy for hyperthermia applications in three frequency bands: 433.05-434.79 MHz (in Europe), 902-928 MHz (in the USA) and 2400-2500 MHz (worldwide) but in medical practice frequencies of 434 MHz, 915 MHz and 2.45 GHz are commonly used [2] as presented in this paper. Because, in the small antennas for interstitial hyperthermia, the whole microwave frequency range (300 MHz – 300 GHz) is limited to a frequency not exceeding 3 GHz, the wavelength varies in the range between 10 cm – 100 cm.

Nowadays, there are many studies on cancer therapy using hyperthermia which demonstrate that this aspect is still important and more research is needed in this matter [6, 7]. More and more scientists are looking for new solutions in the hyperthermia treatment. Recently, the use of nanoparticles in magnetic fluid hyperthermia is under examination [8]. Talking about hyperthermia cannot ignore its historical context [9]. Despite a long history, further investigation into hyperthermia techniques is necessary to make this method simpler, safer, more effective and widely available for patients.

## Main equations and geometrical model

Figures 1 – 2 demonstrate the analysed model of the thin coaxial-slot antenna. It includes such elements as central conductor, dielectric, outer conductor and a plastic catheter, which performs the protective function for all other elements of the antenna. The air gap with size $d$ is located in the outer conductor. The antenna dimensions are taken from [10] and summarized in Table 1. In fact, the computational area is much larger than it is presented in Fig. 1, and the antenna width does not exceed 2 mm. Due to the axial symmetry of the model cylindrical coordinates $r$, $\phi$, $z$ are used. The 2D model, which includes only half of the antenna structure and the surrounding human brain tissue, is sufficient for analysis. The whole 3D model can be obtained by rotating the 2D model along the $z$ axis.

To derive the basic formulas let us start with the Maxwell's equations in the time domain as following:

(1) $$\nabla \times \mathbf{H} = \mathbf{J} + \frac{\partial \mathbf{D}}{\partial t}$$

(2) $$\nabla \times \mathbf{E} = -\frac{\partial \mathbf{B}}{\partial t}$$

where **E** and **H** are the electric and magnetic field strengths, respectively, **J** is the current density and **D** and **B** are adequately the vectors of electric displacement density and magnetic induction.

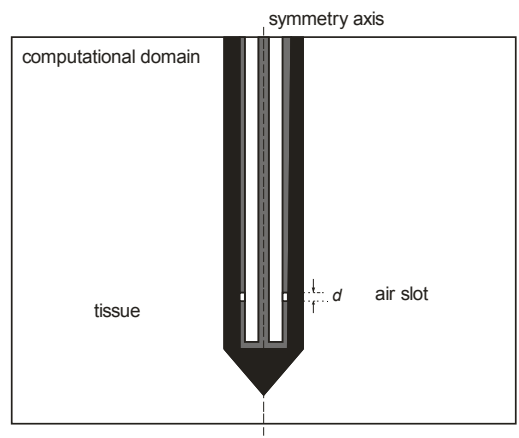

Fig.1. Model of the coaxial antenna located in the human tissue

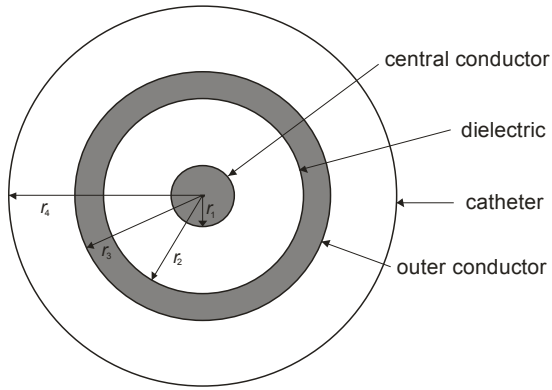

Fig.2. Cross section of the antenna with geometrical dimensions

Table 1. Geometrical dimensions of the antenna in mm

| | |
|---|---|
| radius of the central conductor | $r_1 = 0.135$ |
| inner radius of the outer conductor | $r_2 = 0.470$ |
| outer radius of the outer conductor | $r_3 = 0.595$ |
| radius of the catheter | $r_4 = 0.895$ |
| size of the air slot | $d = 1$ |

In the frequency domain the time derivatives are replaced by the factor j$\omega$ and all the vectors are changed by their complex equivalents according to the equation

$$(3) \qquad \mathbf{H}(\mathbf{r},t) = \mathrm{Re}\left[\mathbf{H}(\mathbf{r})e^{j\omega t}\right]$$

where j is the imaginary unit and $\omega$ is the angular frequency of the electromagnetic field.

After taking into account the Ohm's law in differential form $\mathbf{J} = \sigma \mathbf{E}$ and material dependences $\mathbf{D} = \varepsilon_0 \varepsilon_r \mathbf{E}$ and $\mathbf{B} = \mu_0 \mu_r \mathbf{H}$, for uniform, linear and isotropic medium, equations (1) and (2) in the complex domain take the form

$$(4) \qquad \nabla \times \mathbf{H} = j\omega\varepsilon_0 \left(\varepsilon_r - j\frac{\sigma}{\omega\varepsilon_0}\right)\mathbf{E}$$

$$(5) \qquad \nabla \times \mathbf{E} = -j\omega\mu_0\mu_r\mathbf{H}$$

where $\varepsilon_0$ and $\mu_0$ are the permittivity and permeability of the vacuum, $\varepsilon_r$ and $\mu_r$ are the relative permittivity and relative permeability of the medium, respectively and where $\sigma$ is the electrical conductivity of the body.

After applying the rotation operator to both sides of equation (4) and substituting equation (5) into such converted equation (4) the following equation can be derived, describing the field distribution in the computational domain

$$(6) \qquad \nabla \times \left[\underline{\varepsilon}_r^{-1} \nabla \times \mathbf{H}\right] - k_0^2 \mu_r \mathbf{H} = 0$$

where $\underline{\varepsilon}_r$ is the complex relative permittivity specified by

$$(7) \qquad \underline{\varepsilon}_r(\omega) = \varepsilon_r - j\frac{\sigma}{\omega\varepsilon_0}$$

Moreover, $k_0$ is the free-space wave number defined as

$$(8) \qquad k_0 = \frac{\omega}{c_0} = \omega\sqrt{\varepsilon_0\mu_0}$$

and $c_0$ is the speed of the light in vacuum.

In the analysed model transverse magnetic (TM) waves are used and there are no electric field variations in the azimuthal direction. A magnetic field has only the $\phi$ – component and an electric field propagates in the r-z plane, which can be written as:

$$(9) \qquad \mathbf{H} = H_\phi \mathbf{e}_\phi$$

$$(10) \qquad \mathbf{E} = E_r \mathbf{e}_r + E_z \mathbf{e}_z$$

In the case of an axial-symmetrical model, the wave equation takes the form of scalar equation as bellow

$$(11) \qquad \nabla \times \left[\left(\varepsilon_r - j\frac{\sigma}{\omega\varepsilon_0}\right)^{-1} \nabla \times H_\phi\right] - k_0^2 \mu_r H_\phi = 0$$

The presented problem assumes that the all metallic surfaces using PEC (perfect electric conductor) boundary conditions are modelled

$$(12) \qquad \mathbf{n} \times \mathbf{E} = 0$$

where $\mathbf{n}$ is the unit vector normal to the surface. Moreover, the external boundaries of the computational domain, which do not represent a physical boundary (except the axis of symmetry z) have the so-called matched boundary conditions that make the boundaries totally non-reflecting. This boundary conditions assume the form

$$(13) \qquad \sqrt{\varepsilon - j\frac{\sigma}{\omega}}\,\mathbf{n} \times \mathbf{E} - \sqrt{\mu}H_\phi = -2\sqrt{\mu}H_{\phi 0}$$

where $H_{\phi 0}$ is an input field incident on the antenna like this

$$(14) \qquad H_{\phi 0} = \frac{1}{Zr}\sqrt{\frac{ZP_{in}}{\pi\ln(r_2/r_1)}}$$

In the equation above $P_{in}$ is the total input power in the dielectric, while $r_1$ and $r_2$ are the dielectric's inner and outer radii, respectively. Further $Z$ signifies the wave impedance of the dielectric, which is given by the formula

$$(15) \qquad Z = \frac{Z_0}{\sqrt{\varepsilon_r}} = \frac{120\pi}{\sqrt{\varepsilon_r}}$$

where $Z_0$ is the wave impedance in vacuum.

Table 2. Electro-thermal parameters of the tissues for various microwave frequencies used in the medical practice [11, 12]

| Frequency | $f = 434$ MHz | | $f = 915$ MHz | | $f = 2450$ MHz | | $k$ (W m$^{-1}$ K$^{-1}$) |
|---|---|---|---|---|---|---|---|
| Tissue | $\varepsilon_r$ | $\sigma$ (S m$^{-1}$) | $\varepsilon_r$ | $\sigma$ (S m$^{-1}$) | $\varepsilon_r$ | $\sigma$ (S m$^{-1}$) | |
| brain | 55.11 | 1.048 | 49.35 | 1.270 | 44.80 | 2.101 | 0.51 |
| breast | 61.33 | 0.886 | 59.65 | 1.044 | 57.20 | 1.968 | 0.33 |
| kidney | 65.43 | 1.110 | 58.56 | 1.401 | 52.74 | 2.430 | 0.53 |
| liver | 50.67 | 0.668 | 43.76 | 0.861 | 43.03 | 1.686 | 0.52 |
| lung | 23.58 | 0.380 | 21.97 | 0.459 | 20.48 | 0.842 | 0.39 |

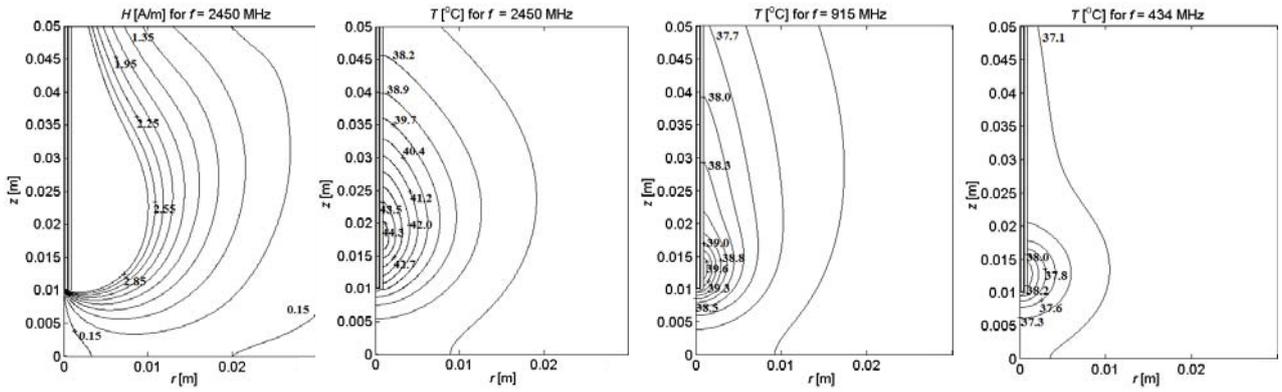

Fig.3. Equipotential lines of the module of the magnetic field strength for the antenna operating frequency of 2.45 GHz and temperature distributions for frequencies 2.45 GHz, 915 MHz and 434 MHz, respectively in order from left to right (all for the kidney tissue and $P_{in} = 1$W)

Furthermore, on the symmetry axis of the antenna it is assumed that

$$E_r = 0, \quad \frac{\partial E_z}{\partial r} = 0 \quad (16)$$

The seed point is modelled using a port boundary condition with the power level set to $P_{in} = 1$W at the low-reflection external boundary of the coaxial dielectric cable. The full derivation of wave equation (11) can be found in [13].
 The electromagnetic field is coupled with temperature field using the so-called bioheat equation given by Pennes in the mid-twentieth century [14]. The Pennes equation describes the phenomenon of heat transfer in biological tissues and in steady-state analysis it is expressed by

$$\nabla(-k\nabla T) = \rho_b C_b \omega_b (T_b - T) + Q_{ext} + Q_{met} \quad (17)$$

where $T$ is the body temperature (K), $k$ – the tissue thermal conductivity (W m$^{-2}$ K$^{-1}$), $T_b$ – the blood vessel temperature (K), $\rho_b$ – the blood density (kg m$^{-3}$), $\omega_b$ – the blood perfusion rate (s$^{-1}$), $C_b$ – the blood specific heat (J kg$^{-1}$ K$^{-1}$). What is interesting, the described model takes into account the so-called metabolic heat generation $Q_{met}$ (W m$^{-3}$) as well as the external heat sources $Q_{ext}$ (W m$^{-3}$), which are relevant to power densities. The last one is responsible for the changing of the temperature inside human tissue as

$$Q_{ext} = 0.5\, \sigma\, \mathbf{E} \cdot \mathbf{E}^* = 0.5\, \sigma |\mathbf{E}|^2 \quad (18)$$

Since the computational domain is limited to a part of the human tissue, it can be assumed that the heat exchange between parts of the same tissue does not occur and the boundary condition describing this process uses thermal insulation given by

$$\mathbf{n} \cdot (k\nabla T) = 0 \quad (19)$$

**Simulation results**
 In the analyzed model, the human tissue and antenna are considered as uniform mediums with averaged material parameters. The antenna operates at different frequencies designated by ITU. Since the human tissues are the dispersion media, therefore their electrical parameters depend on the frequency, as shown in Table 2 for different tissues. The blood parameters are given in Table 3. Other electrical parameters of the antenna are put together in Table 4. Moreover, the metabolic heat generation of the tissue is neglected because usually $Q_{met} \ll Q_{ext}$. Both main equations (11) and (17) with the appropriate boundary conditions method were solved using the finite element.

Table 3. Physical parameters of blood taken in the model [12]

| Tissue | $\rho_b$ (kg m$^{-3}$) | $C_b$ (J kg$^{-1}$·K$^{-1}$) | $T_b$ (K) | $\omega_b$ (s$^{-1}$) |
|---|---|---|---|---|
| Blood | 1020 | 3640 | 310.15 | 0.004 |

Table 4. Electrical parameters of the coaxial-slot antenna [10]

| | $\varepsilon_r$ | $\mu_r$ | $\sigma$ (S m$^{-1}$) |
|---|---|---|---|
| dielectric | 2.03 | 1 | 0 |
| catheter | 2.60 | 1 | 0 |
| air slot | 1 | 1 | 1 |

All simulation results that are summarized in Figures 3 – 7 have been calculated for $P_{in} = 1$ W. Fig. 3 represents the distribution of the module of the magnetic field (for 2.45 GHz) and isotherms (for three values of the frequency) in the kidney tissue. Temperature distributions along two paths crossing the kidney tissue at $z = 16$ mm (height of the air slot) and $r = 2.5$ mm for various frequencies are presented in next two Figures 4 – 5.

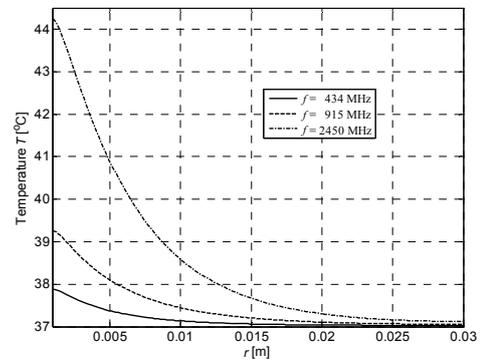

Fig.4. Temperature distributions for the kidney tissue along the path $z = 16$ mm for different frequencies and $P_{in} = 1$W

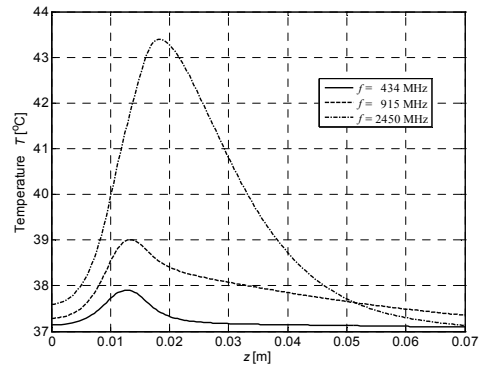

Fig.5. Temperature distributions for the kidney tissue along the path $r = 2.5$ mm for different frequencies and $P_{in} = 1$W

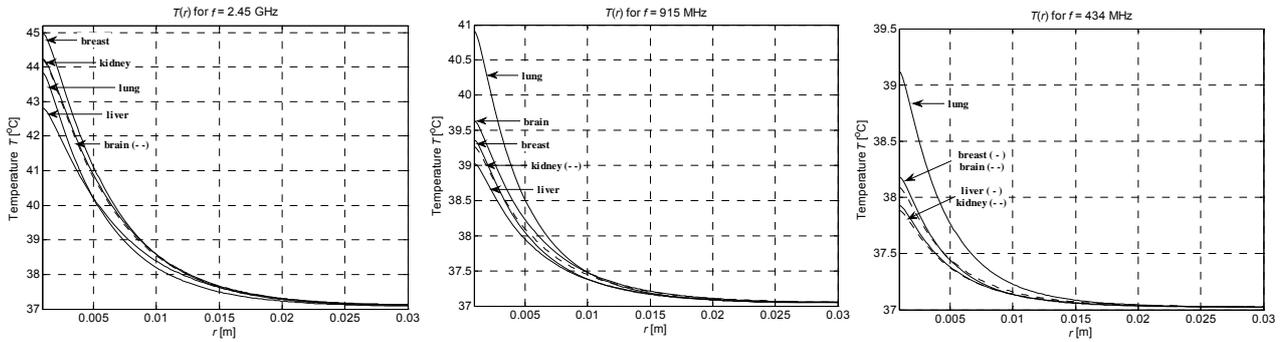

Fig.6. Temperature distribution for various tissues along the path $z = 16$ mm for different frequencies and $P_{in} = 1$ W

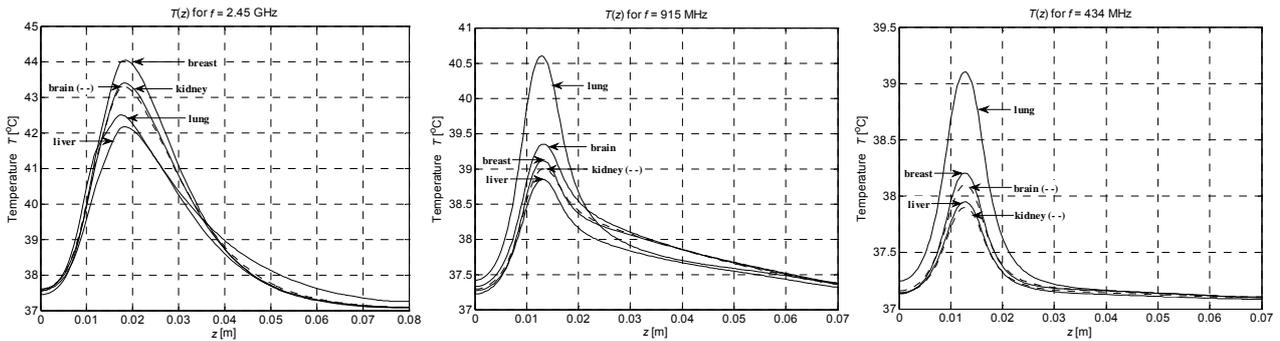

Fig.7. Temperature distributions for various tissues along the path $r = 2.5$ mm for different frequencies and $P_{in} = 1$ W

Next illustrations (Figures 6 – 7) show the temperature distributions in different human tissues along two specified paths for different frequencies used in medical practise.

**Summary**

Interstitial microwave hyperthermia is an invasive type of thermal therapy, in which heat produced by microwaves is used to kill pathological cells associated with tumors located deep with the human body. Because this special technique is now gaining new fields of application in the treatment of liver, breast, kidney, brain and lung tumors, these types of tissues were used in the simulations. It should be remembered that tissue parameters such as permittivity as well as electric and thermal conductivity are closely connected with the frequency and type of human tissue. As expected, the temperature inside tissue decreases rapidly with the distance from the microwave antenna and its largest values occur near the antenna's air gap. For frequency of $2.45$ GHz, the greatest and the lowest temperatures occur in breast and liver, respectively. Interestingly, very similar temperature values can be observed in the kidney and brain tissues. For lower frequencies the tissue set changes as a direct result of their electro-thermal parameters. Now the highest temperatures appear in the lung tissue and the smallest in the kidney and liver tissue. What is more, the temperature induced in the tissue is reduced. The values of the tissue temperature can be easily enlarged by increasing the antenna's input power.

REFERENCES

[1] Habash R.W., Krewski D., Bansal R., Alhafid H.T., Principles, Applications, Risks and Benefits of Therapeutic Hyperthermia, *Frontiers in Bioscience* (*Elite Edition*), 3 (2011), 1169-1181.
[2] Dębicki P., Hipertermia mikrofalowa w leczeniu chorób nowotworowych – problem planowania leczenia i sterowania zabiegiem, *Pomiary Automatyka Kontrola*, vol. 49 (2003), nr 1, str. 45-48.
[3] Rubio M.F.J.C., Hernandez A.V., Salas L.L., Ávila-Navarro E., Navarro E.A., Coaxial Slot Antenna Design for Microwave Hyperthermia using Finite-Difference Time-Domain and Finite Element Method, *Open Nanomedicine Journal*, 3 (2011), SPEC. ISSUE, 2-9.
[4] Baronzio G.F., Hager E.D., *Hyperthermia in Cancer Treatment: A Primer*, Landes Bioscience and Springer Science + Business Media, New York (2006).
[5] Report ITU-R SM.2180: *Impact of Industrial, Scientific and Medical (ISM) Equipment on Radiocommunication Services*, ITU, Geneva (2011).
[6] Gas P., Temperature inside Tumor as Time Function in RF Hyperthermia, *Electrical Review*, 86 (2010), No. 12, 42-45.
[7] Kurgan E., Gas P., Distribution of the Temperature in Human Body in RF Hyperthermia, *Electrical Review*, 85 (2009), No. 1, 96-99.
[8] Miaskowski A., Krawczyk A., Magnetic Fluid Hyperthermia for Cancer Therapy, *Electrical Review*, 87 (2011), No. 12b, 125-127.
[9] Gas P., Essential Facts on the History of Hyperthermia and their Connections with Electromedicine, *Electrical Review*, 87 (2011), No. 12b, 37-40.
[10] Saito K., Hosaka S., Okabe S.Y., A proposition on improvement of a heating pattern of an antenna for microwave coagulation therapy: introduction of a coaxial-dipole antenna, *Electronics and Communications in Japan (Part I: Communications)*, 86 (2003), No. 1, 16-23.
[11] Gabriel S., Lau R.W., Gabriel C., The dielectric properties of biological tissues: III. Parametric models for the dielectric spectrum of tissues, *Phys. Med. Biol.*, 41 (1996), 2271-2293.
[12] Mcintosh R. L., Anderson V., A Comprehensive Tissue Properties Database Provided for the Thermal Assessment of a Human at Rest, *Biophysical Reviews and Letters*, 5 (2010), No. 3, 129-151.
[13] Gas P., Temperature Distribution of Human Tissue in Interstitial Microwave Hyperthermia, *Electrical Review*, Vol. 88 (2012), No. 7a, 144-146.
[14] Pennes H.H., Analysis of Tissue and Arterial Blood Temperatures in the Resting Human Forearm, *J. Appl. Physiol.*, 1 (1998), No. 85, 5-34.

*Authors*: mgr inż. Piotr Gas, AGH University of Science and Technology, Department of Electrical and Power Engineering, al. Mickiewicza 30, 30-059 Kraków, E-mail: piotr.gas@agh.edu.pl